%% file: NoSPaM.tex
\documentclass[12pt,a4paper,abstract]{scrartcl}
\usepackage[numbers, square, comma, sort&compress]{natbib}

\usepackage{amsmath}           
\usepackage{amsfonts}          
\usepackage{amssymb}           
\usepackage{lmodern,dsfont}	
\usepackage{color}
\usepackage{graphicx}          
\usepackage[T1]{fontenc}       

\usepackage[margin=10pt,font=small,labelfont=bf,format=plain,indention=0.cm]{caption} 
\usepackage{subfig}
\usepackage{scrhack} 
\usepackage{lmodern,dsfont}	
\usepackage{algorithm}
\usepackage{algpseudocode}
\usepackage{algorithmicx}
\usepackage[breaklinks,hidelinks]{hyperref}
\usepackage{wrapfig}

\input{T_motifSymbols.tex}

\newcommand{\qed}{\nobreak \ifvmode \relax \else
      \ifdim\lastskip<1.5em \hskip-\lastskip
      \hskip1.5em plus0em minus0.5em \fi \nobreak
      \vrule height0.75em width0.5em depth0.25em\fi}

\begin{document}%

\title{NoSPaM Manual}
\subtitle{A Tool for Node-Specific Triad Pattern Mining}
\author{Marco Winkler\thanks{Institute for Theoretical Physics, University of W\"{u}rzburg, Am Hubland, 97074 W\"{u}rzburg, Germany, mwinkler@physik.uni-wuerzburg.de}}

	
\date{\today}

\maketitle

\begin{abstract}

The detection of triadic subgraph motifs is a common methodology in complex-networks research. The procedure usually applied in order to detect motifs evaluates whether a certain subgraph pattern is overrepresented in a network as a whole. However, motifs do not necessarily appear frequently in every region of a graph. For this reason, we recently introduced the framework of \textbf{No}de-\textbf{S}pecific \textbf{Pa}ttern \textbf{M}ining~(\textsc{NoSPaM})~\cite{Winkler2015diss, Winkler2014}. This work is a manual for an implementation of \textsc{NoSPaM} which can be downloaded from \url{www.mwinkler.eu}.

\end{abstract}

\tableofcontents

\newpage


\section{Introduction}

\begin{figure}
	\centering
		\includegraphics[width=0.935\textwidth]{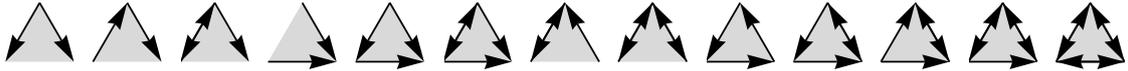}
	\caption{All 13 possible non-isomorphic connected triadic subgraphs (subgraph patterns) in directed unweighted networks.}
	\label{fig:triadPatterns}
\end{figure}

Analyzing networks in terms of their local substructure is a well-established methodology in complex-network science~\cite{Milo2002,Milo2004,Sporns2004,Albert2004,berlingerio2009mining,rahman2014graft}. Particularly triadic subgraph structures have been studied intensively over the last 15 years~\cite{Milo2002,Milo2004,Alon2007,Winkler2013}. Apart from node permutations, there are 13 connected triadic subgraph patterns in directed unweighted networks~(see Fig.~\ref{fig:triadPatterns}). Those patterns that are significantly overrepresented in a graph structure are referred to as \textit{motifs}~\cite{Milo2002}.

In order to evaluate the extent of an over- or underrepresentation, for each pattern $i$, the framework commonly used compares its frequency of occurrence in the original network under investigation, $N_{\text{original},i}$, to the expected frequency of occurrence in an ensemble of random networks with the same degree distribution and the same number of unidirectional and bidirectional links as the original network, $\left\langle N_{\text{rand},i}\right\rangle$. Over- and underrepresentation of pattern $i$ is then quantified through a $Z$~score
\begin{equation}
	Z_{i} = \frac{N_{\text{original},i} - \left\langle N_{\text{rand},i}\right\rangle}{\sigma_{\text{rand},i}}
	\label{eq:Zscore}
\end{equation}
where $\sigma_{\text{rand},i}$ represents the standard deviation of $ N_{\text{rand},i}$ in the ensemble of the null model. Hence, every network can be assigned a vector $\vec{Z}$ whose components comprise the $Z$ scores of all possible triad patterns of Fig.~\ref{fig:triadPatterns}.

\begin{figure}
	\centering
	\includegraphics[width=0.980\textwidth]{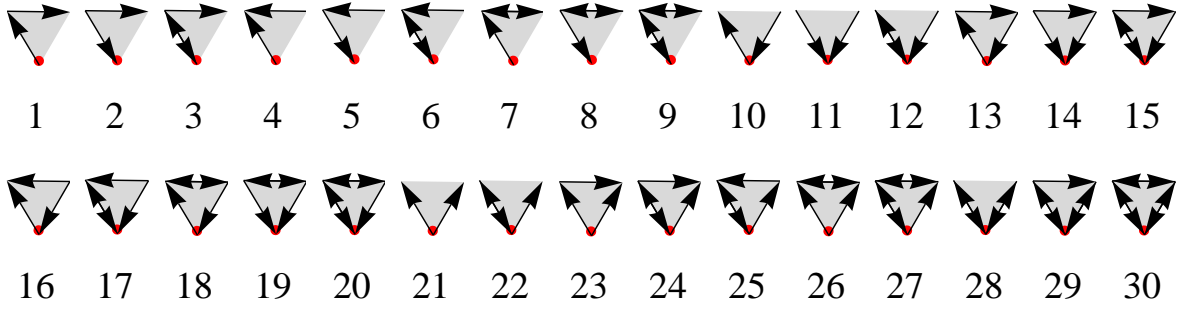}
	\caption{All possible connected, nonisomorphic triadic subgraph patterns in terms of a distinct node (here: lower node).}
	\label{fig:dir_triadLocalPatterns_connected}
\end{figure}

However, this approach does not account for potentially existing heterogeneities in graph structures. Suppose, e.g., the feed-forward loop (FFL), $\motVIII$, is overrepresented in a certain area of a graph, but, at the same time, highly underrepresented in another part of the graph. On the global network level, the effects may cancel out such that the $Z$ score will be close to zero and possibly relevant structural information will be lost. Therefore, we recently suggested the methodology of \textbf{No}de-\textbf{S}pecific \textbf{Pa}ttern \textbf{M}ining~(\textsc{NoSPaM})~\cite{Winkler2015diss, Winkler2014}. Instead of mining frequent subgraphs on the system level, \textsc{NoSPaM} investigates the neighborhood of every single node separately. I.e., for every node $\alpha$, \textsc{NoSPaM} considers only those triads in which $\alpha$ participates in. Since the position of node $\alpha$ in the triadic subgraphs matters now, the symmetry of most patterns shown in Fig.~\ref{fig:triadPatterns} is broken and the number of connected \textit{node-specific} triad patterns increases from 13 to 30. These are shown in Fig.~\ref{fig:dir_triadLocalPatterns_connected}. To understand the increase in the number of patterns, consider the ordinary subgraph~$\motVIII$. From the perspective of one particular node, it splits into the three node-specific triad patterns 14, 16, and~23 in Fig.~\ref{fig:dir_triadLocalPatterns_connected}. Furthermore, some patterns are included in others, e.g. pattern~1 is a subset of pattern~3. In order to avoid biased results, it is not double counted, i.e. an observation of pattern~3 will only increase its corresponding count and not the one associated with pattern~1~\cite{Winkler2015diss,Winkler2014}.

For every node $\alpha$ in a graph, \textsc{NoSPaM} will compute $Z$~scores for each of the 30 node-specific patterns $i$ shown in Fig.~\ref{fig:dir_triadLocalPatterns_connected},
\begin{equation}
	Z_{i}^{\alpha} = \frac{N_{\text{original},i}^{\alpha} - \left\langle N_{\text{rand},i}^{\alpha}\right\rangle}{\sigma_{\text{rand},i}^{\alpha}}.
	\label{eq:nodeZscore}
\end{equation}
$N_{\text{original},i}^{\alpha}$ is the number of appearances of pattern $i$ in the triads node $\alpha$ participates in. Accordingly, $\left\langle N_{\text{rand},i}^{\alpha}\right\rangle$ is the expected frequency of pattern $i$ in the triads node $\alpha$ is part of in the ensemble of graphs with the same degree distribution and the same number of both unidirectional and bidirectional links. $\sigma_{\text{rand},i}^{\alpha}$ is the corresponding standard deviation.

The following section of this manual provides for information on how to download and run an implementation of \textsc{NoSPaM}. In Section~\ref{sec:alg}, details of the applied algorithms are discussed. 


\newpage

\section{How to use NoSPaM}

\subsection{Download and Installation}

\begin{itemize}
	\item \textbf{Requirements:} make sure to have a recent Java version installed (version 1.7 or higher) 
	\item \textbf{Download:} \textsc{NoSPaM} can be downloaded from \url{www.mwinkler.eu} following the \textit{Supplementary Material} link.
	\item Extract the \verb nospam.zip ~file to whereever you prefer to store \textsc{NoSPaM}.
	\item Go to the command line terminal and navigate to the nospam directory.
	\item Generate executable class files by typing: \verb"javac *.java"
	\item $^*$In case of problems with the compilation process make sure that the PATH variable includes the JDK, e.g. \verb"'C:\Program Files\Java\jdk1.7.0_11\bin'"
	\item Start a test run by typing: \verb"java nospam exampleNetwork.txt 2000 1500"
\end{itemize}

\subsection{Running NoSPaM} 

Input data must be stored in the followig format:
\begin{verbatim}
	<source node 1><tab><target node 1>
	<source node 2><tab><target node 2>
	...
	<source node M><tab><target node M>
\end{verbatim}
Every line represents an edge of the network. The identities of the node must be represented by integer values. Any additional entries in a line (e.g. an edge weight) will be ignored by the algorithm. The file type is arbitrary and can be, e.g., \verb".txt", \verb".dat", \verb".csv", etc...

The general command line syntax is as follows:
\begin{verbatim}
	java nospam <filePath> <samples> <switching attempts>
\end{verbatim}

\begin{itemize}
	\item if the network file to be analyzed is already in the \verb"nospam" directory (e.g. the file \verb"exampleNetwork.txt"), the variable \verb"<filePath>" can simply be set to the value \verb"<fileName>.<fileType>"
	\item the variable \verb"<samples>" specifies the number of instances from the randomized ensemble to be used for estimating the average values, $\left\langle N_{\text{rand},i}^{\alpha}\right\rangle$, and the standard deviations, $\sigma_{\text{rand},i}^{\alpha}$.
	\item the variable \verb"<switching attempts>" specifies the number of microscopic rewiring steps. It should be chosen proportionally to the number of edges, $|E|$, in the graph and not smaller than $|E|$ (see Section~\ref{sec:alg} and References~\cite{milo2003uniform} and~\cite{Winkler2015diss} for details).
\end{itemize}

\subsection{Output File}

\begin{figure}
	\centering
		\includegraphics[width=1.00\textwidth]{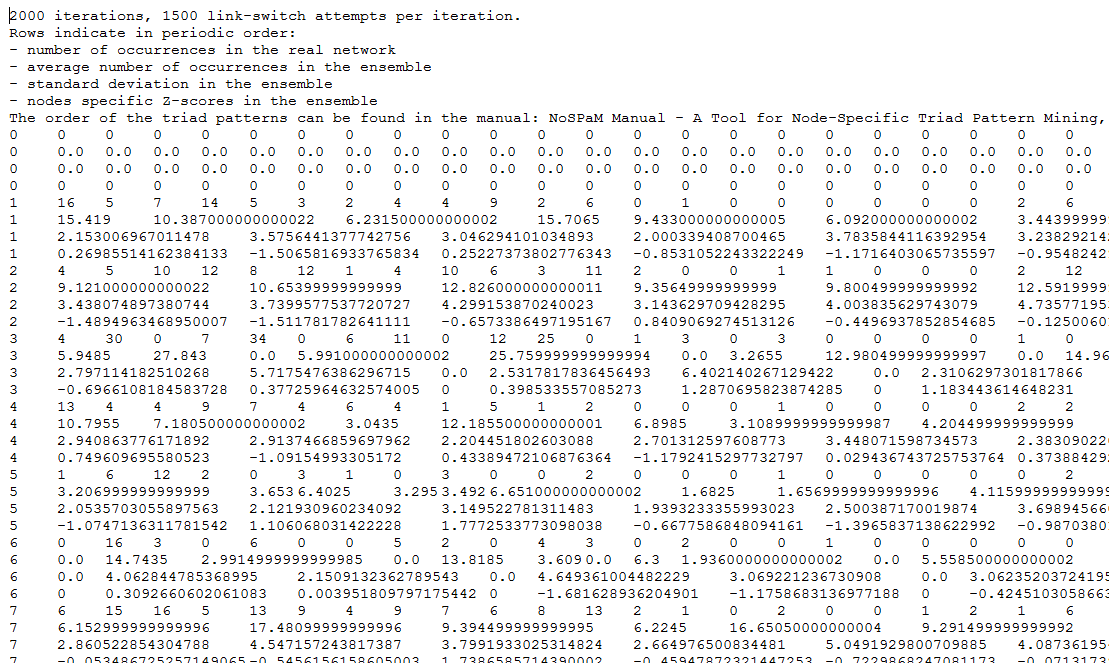}
	\caption{Example output file. The first column indicates the ids of the nodes. The following 30 columns correspond to the node-specific patterns shown in Fig.~\ref{fig:dir_triadLocalPatterns_connected} (in the same order). For every node, there are four lines: the first one indicates the total number of occurrences in the original network, $N_{\text{original},i}^{\alpha}$; the second line indicates the mean number of occurrences in the randomized ensemble, $\left\langle N_{\text{rand},i}^{\alpha}\right\rangle$; the third line indicates the standard deviation in the randomized ensemble, $\sigma_{\text{rand},i}^{\alpha}$; the fourth line indicates the node-specific $Z$ scores which can be obtained from the other three measures.}
	\label{fig:output}
\end{figure}

The output of \textsc{NoSPaM} is written in a separate file of the same type as the input file. Fig.~\ref{fig:output} gives an example of such an output file. After the header lines, the evaluated data is stored. For each node (indicated by the ids in the first column), the values of $N_{\text{original},i}^{\alpha}$, $\left\langle N_{\text{rand},i}^{\alpha}\right\rangle$, $\sigma_{\text{rand},i}^{\alpha}$, and $Z_{i}^{\alpha}$ are stored in the columns corresponding to the patterns $i$ in Fig.~\ref{fig:dir_triadLocalPatterns_connected}.

For the output file shown in Fig.~\ref{fig:output}, e.g., from the perspective of the node with id 1, pattern 1 ($\nmotI$) has appeared 16 times in the network under investigation. In the randomized ensemble it appeared 15.42 times on average with a standard deviation of $2.15$ resulting in a $Z$ score, $Z_{i=1}^{\alpha=1}$ of $0.27$ indicating a non-significant overrepresentation of the pattern. Further, pattern 2 ($\nmotII$) has appeard 5 times in the neighborhood of node 1, while in the randomized ensemble it occurred $10.39$ times on average with a standard deviation of $3.58$ yielding a $Z$ score of $Z_{i=2}^{\alpha=1} = -1.51$ indicating an underrepresentation of the pattern.


\newpage

\section{Algorithms} \label{sec:alg}

\subsection{Randomization}

\begin{figure}
	\centering
	\subfloat[][]{
		\centering
		\includegraphics[width=0.45\textwidth]{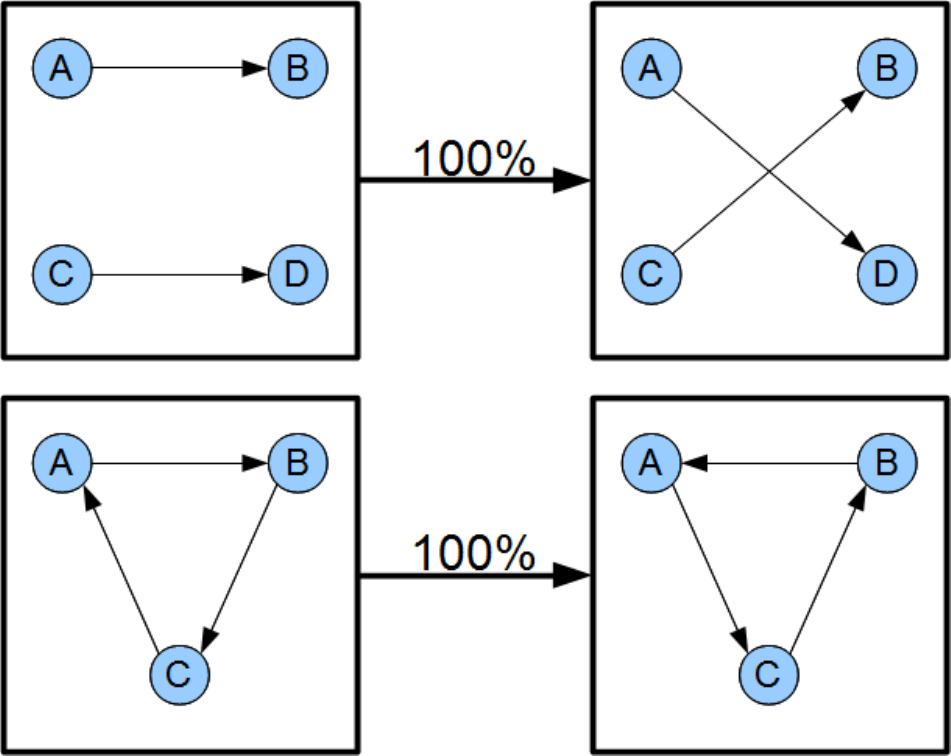}
	\label{fig:randomizationUni}
	}
	\hspace{0.1cm}
	\subfloat[][]{
		\includegraphics[width=0.45\textwidth]{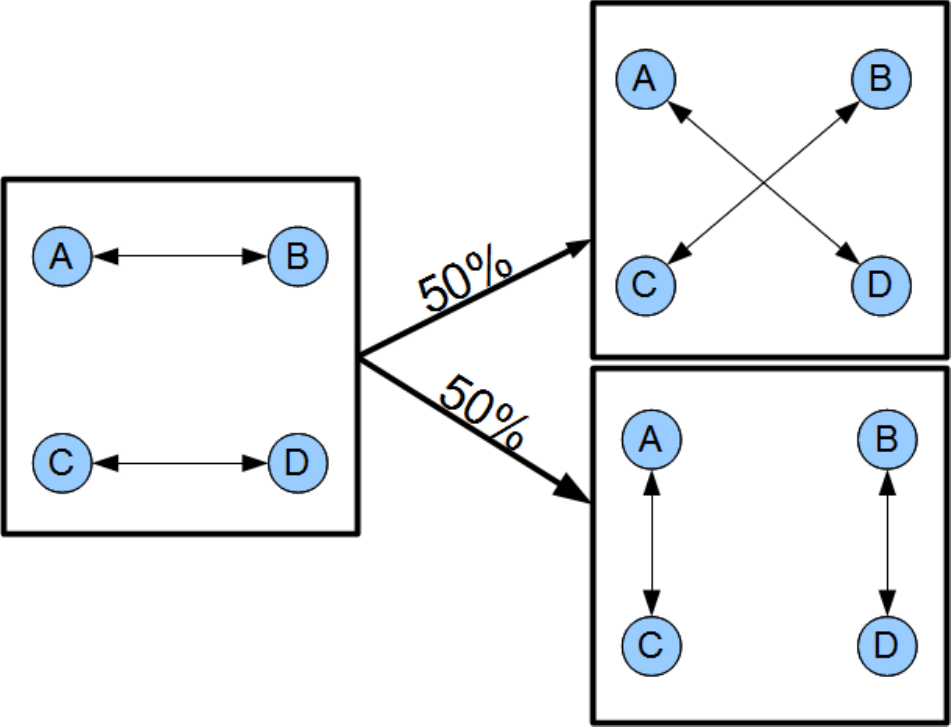}
	\label{fig:randomizationBi}
	}
	\caption{Microscopic link-switchings performed to generate the randomized ensembles. \textbf{(a)}~\textit{Pair switch} and \textit{loop switch} for unidirectional links. \textbf{(b)}~\textit{Pair switch} for bidirectional links.}
	\label{fig:randomizationRules}
\end{figure}

The generation of the random null model with the same degree distribution and the same number of unidirectional and bidirectional links adjacent to each vertex as in the network under investigation is realized by a link-swapping algorithm. The microscopic switching rules are illustrated in Fig.~\ref{fig:randomizationRules}.

The number of microscopic switching attempts is the only parameter of the randomization that needs to be specified in advance. The entire randomization procedure is displayed in Algorithm~\ref{alg:randomization}. The Markov chain generated by successive link swappings obeys detailed balance and coverges towards a uniform distribution of the networks in the ensemble serving as the null model, i.e. every valid network is sampled with equal probability. This fact is elaborated in depth in~\cite{Winkler2015diss}. The number of switching attempts should be chosen proportionally to the number of edges in the graph~\cite{milo2003uniform}.

\begin{algorithm}
  \caption{Degree-preserving randomization of a graph
    \label{alg:randomization}}
  \begin{algorithmic}
	\Function{Randomize}{Graph $\mathcal{G}\left(V,E\right)$, no. of required steps}
    \State s = 0
    \While{s $<$ number of required rewiring steps}
    \State pick a random link $e_1 \in E$
    \If{$e_1$ is unidirectional}
    	\State pick a 2nd unidirectional link $e_2 \in E$ at random
    \Else
    	\State pick a 2nd bidirectional link $e_2 \in E$ at random
    \EndIf
    \State
    \If{$e_1$ and $e_2$ do not share a node}
    \State rewire according to the pair-switch rules in Fig.~\ref{fig:randomizationRules}
    	\If{one of the new links already exists}
    		\State undo the rewiring
    	\EndIf
    \ElsIf{$e_1$ and $e_2$ participate in a loop}
    \State rewire according to the loop-switch rule in Fig.\ref{fig:randomizationUni}
    \EndIf
    \State
    \State s++
    \EndWhile
    \State \Return randomized instance of $\mathcal{G}$
   \EndFunction
  \end{algorithmic}
\end{algorithm}

\subsection{Counting of Node-Specific Triad Patterns}

For counting the frequencies in which the distinct node-specific triad patterns occur, an iteration over all connected triads is necessary. The procedure is illustrated in Algorithm~\ref{alg:nspTriadCounting}. Because it is computationally expensive to test all triads in the system (the complexity is of order $\mathcal{O}\left(N^3\right)$), we rather iterate over pairs of adjacent edges in the graph. Since real-world networks are usually sparse, this is much more efficient~\cite{Winkler2015diss}.

\begin{algorithm}
  \caption{Counting of node-specific triad patterns
    \label{alg:nspTriadCounting}}
  \begin{algorithmic}
	\Function{NspPatternCounter}{Graph $\mathcal{G}(V,E)$}
		\State $\mathcal{N}$: $N \times 30$-dimensional array storing the pattern counts for every node of $\mathcal{G}$
    \For{every edge $e \in E$}
    	\State $i, j$ $\leftarrow$ IDs of $e$'s nodes with $i < j$
    	\State $\mathcal{C} \leftarrow \left\{\right\}$ be list of candidate nodes to form triad patterns comprising $e$
    	\State $\mathcal{C} \leftarrow $ all neighbors of $i$
    	\State $\mathcal{C} \leftarrow $ all neighbors of $j$
    	\For{all $c \in \mathcal{C}$}
    		\If{$i+j < $ sum of IDs of all other \textit{connected}\\ \qquad \qquad \qquad \qquad \qquad dyads in triad $(ijc)$}
    			\State increase the counts in $\mathcal{N}$ for $i$, $j$, and $c$ for 
    			\State their respective node-specific patterns
    		\EndIf
    	\EndFor
    \EndFor
    \State \Return $\mathcal{N}$
   \EndFunction
  \end{algorithmic}
\end{algorithm}

\subsection{Node-Specific Triad Pattern Mining}

The evaluation of the node-specific $Z$ scores is performed by Algorithm~\ref{alg:nospam3}. For a more detailed discussion of the algorithms including considerations of the compuational effort see References~\cite{Winkler2014,Winkler2015diss}.

\begin{algorithm}
  \caption{Node-specific triad pattern mining (\textsc{NoSPaM})
    \label{alg:nospam3}}
  \begin{algorithmic}
	\Function{NoSPaM}{Graph $\mathcal{G}$, \# required rewiring steps, \#~randomized instances}
		\State $\mathcal{N}_\text{original} \leftarrow$ \textsc{NspPatternCounter}($\mathcal{G}$)
		\State $\mathcal{N}_\text{rand} \leftarrow \left\{\right\}$ 
		\State $\mathcal{N}_\text{sq,rand} \leftarrow \left\{\right\}$
		\For{\# randomized instances}
			\State $\mathcal{G} \leftarrow $ \textsc{Randomize}($\mathcal{G}$, \# required rewiring steps)
			\State counts $\leftarrow$ \textsc{NspPatternCounter}($\mathcal{G}$) 
			\State $\mathcal{N}_\text{rand} \leftarrow \mathcal{N}_\text{rand} +$ counts
			\State $\mathcal{N}_\text{sq,rand} \leftarrow \mathcal{N}_\text{sq,rand} +$ counts $*$ counts
		\EndFor
		\State $\mathcal{N}_\text{rand} \leftarrow \mathcal{N}_\text{rand} / (\# \text{randomized instances})$ 
		\State $\mathcal{N}_\text{sq,rand} \leftarrow \mathcal{N}_\text{sq,rand} / (\# \text{randomized instances})$ 
		\State $\mathcal{\sigma}_\text{rand} \leftarrow \sqrt{\mathcal{N}_\text{sq,rand}-(\mathcal{N}_\text{rand}*\mathcal{N}_\text{rand})}$ 
		\State $\mathcal{Z} \leftarrow (\mathcal{N}_\text{original} - \mathcal{N}_\text{rand})/\mathcal{\sigma}_\text{rand}$ 
		\State \Return $\mathcal{Z}$
   \EndFunction
  \end{algorithmic}
\end{algorithm}


\newpage

\addcontentsline{toc}{section}{References}

\bibliographystyle{abbrv}
\bibliography{Promotion}{}


%



\end{document}

%% file: T_motifSymbols.tex
\newcommand{\nmotSizeI}{3.1ex}
\newcommand{\nmotSizeII}{2.8ex}
\newcommand{\nmotSizeIII}{2.8ex}
\newcommand{\nmotSizeIV}{0.9ex}
\newcommand{\motSizeI}{3.1ex}
\newcommand{\motSizeII}{2.8ex}
\newcommand{\motSizeIII}{2.8ex}
\newcommand{\motSizeIV}{0.9ex}

\newcommand{\nmotI}{
 {\mathchoice
  {\includegraphics[height=\nmotSizeI]{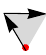}}
  {\includegraphics[height=\nmotSizeII]{figs/nsp_triadConfig_1.pdf}}
  {\includegraphics[height=\nmotSizeIII]{figs/nsp_triadConfig_1.pdf}}
  {\includegraphics[height=\nmotSizeIV]{figs/nsp_triadConfig_1.pdf}}
 }
}
\newcommand{\nmotII}{
 {\mathchoice
  {\includegraphics[height=\nmotSizeI]{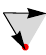}}
  {\includegraphics[height=\nmotSizeII]{figs/nsp_triadConfig_2.pdf}}
  {\includegraphics[height=\nmotSizeIII]{figs/nsp_triadConfig_2.pdf}}
  {\includegraphics[height=\nmotSizeIV]{figs/nsp_triadConfig_2.pdf}}
 }
}

\newcommand{\motVIII}{
 {\mathchoice
  {\includegraphics[height=\motSizeI]{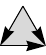}}
  {\includegraphics[height=\motSizeII]{figs/triadConfig_8.pdf}}
  {\includegraphics[height=\motSizeIII]{figs/triadConfig_8.pdf}}
  {\includegraphics[height=\motSizeIV]{figs/triadConfig_8.pdf}}
 }
}

%% file: NoSPaM.bbl
\begin{thebibliography}{10}

\bibitem{Albert2004}
I.~Albert and R.~Albert.
\newblock {Conserved network motifs allow protein-protein interaction
  prediction.}
\newblock {\em Bioinformatics (Oxford, England)}, 20(18):3346--52, Dec. 2004.

\bibitem{Alon2007}
U.~Alon.
\newblock {Network motifs: theory and experimental approaches.}
\newblock {\em Nature reviews. Genetics}, 8(6):450--61, June 2007.

\bibitem{berlingerio2009mining}
M.~Berlingerio, F.~Bonchi, B.~Bringmann, and A.~Gionis.
\newblock Mining graph evolution rules.
\newblock In {\em Machine learning and knowledge discovery in databases}, pages
  115--130. Springer, 2009.

\bibitem{Milo2004}
R.~Milo, S.~Itzkovitz, N.~Kashtan, R.~Levitt, S.~Shen-Orr, I.~Ayzenshtat,
  M.~Sheffer, and U.~Alon.
\newblock {Superfamilies of evolved and designed networks.}
\newblock {\em Science}, 303(5663):1538--42, Mar. 2004.

\bibitem{milo2003uniform}
R.~Milo, N.~Kashtan, S.~Itzkovitz, M.~Newman, and U.~Alon.
\newblock On the uniform generation of random graphs with prescribed degree
  sequences.
\newblock {\em arXiv preprint cond-mat/0312028}, 2004.

\bibitem{Milo2002}
R.~Milo, S.~Shen-Orr, S.~Itzkovitz, N.~Kashtan, D.~Chklovskii, and U.~Alon.
\newblock {Network motifs: simple building blocks of complex networks.}
\newblock {\em Science}, 298(5594):824--7, Oct. 2002.

\bibitem{rahman2014graft}
M.~Rahman, M.~Bhuiyan, and M.~Hasan.
\newblock Graft: An efficient graphlet counting method for large graph
  analysis.
\newblock 2014.

\bibitem{Sporns2004}
O.~Sporns and R.~K\"{o}tter.
\newblock {Motifs in Brain Networks}.
\newblock {\em PLoS Biology}, 2(11):e369, 2004.

\bibitem{Winkler2015diss}
M.~Winkler.
\newblock {\em On the Role of Triadic Substructures in Complex Networks}.
\newblock epubli GmbH, Berlin, 2015.

\bibitem{Winkler2013}
M.~Winkler and J.~Reichardt.
\newblock {Motifs in triadic random graphs based on Steiner triple systems}.
\newblock {\em Phys. Rev. E}, 88(2):022805, Aug. 2013.

\bibitem{Winkler2014}
M.~Winkler and J.~Reichardt.
\newblock Node-specific triad pattern mining for complex-network analysis.
\newblock In {\em IEEE International Conference on Data Mining Workshop
  (ICDMW)}, pages 605--612, Dec 2014.

\end{thebibliography}
